\documentclass[aps,prl,twocolumn,showpacs,superscriptaddress,nofootinbib]{revtex4}

\usepackage{amsmath}   % need for subequations
\usepackage{graphicx}  % for figures
\usepackage{xspace}
\usepackage{units}
\usepackage{lscape}

\usepackage{hyperref}

\newcommand{\minerva}{MINERvA\xspace}

\newcommand{\Eavail}{\ensuremath{E_\mathrm{avail}}\xspace}
\newcommand{\ud}{\ensuremath{\mathrm{d}}}
\newcommand{\dsdEdq}{\ensuremath{\ud^2\sigma/\ud \Eavail \ud q_{3}}\xspace}

\newcommand{\sizecheck}{0} % 0 to do nothing; 1 to check size
\newif\ifpdf
\ifx\pdfoutput\undefined
   \pdffalse
\else
   \pdfoutput=1
   \pdftrue
\fi
\ifpdf
   \usepackage{graphicx}
   \usepackage{epstopdf}
\else
   \usepackage{graphicx}
\fi

\begin{document}

\title{Identification of nuclear effects in neutrino-carbon interactions \\at low three-momentum transfer}


%% MANUAL PARTS OF AUTHOR LIST

%%%%%%RIK  must add Jake, Alec, Miranda, and Move Nuruzzaman.

%% (1) need to add ``\thanks{\deceased}'' after DeMaat, Gobbi, Tzanakos, Zavala
\newcommand{\deceased}{Deceased}

%% AUTOMATIC LIST (EDITED AS ABOVE)
%% List of institution addresses, in command form.

\newcommand{\Rutgers}{Rutgers, The State University of New Jersey, Piscataway, New Jersey 08854, USA}
\newcommand{\Hampton}{Hampton University, Dept. of Physics, Hampton, VA 23668, USA}
\newcommand{\Dortmund}{Institute of Physics, Dortmund University, 44221, Germany }
\newcommand{\Otterbein}{Department of Physics, Otterbein University, 1 South Grove Street, Westerville, OH, 43081 USA}
\newcommand{\JMU}{James Madison University, Harrisonburg, Virginia 22807, USA}
\newcommand{\Florida}{University of Florida, Department of Physics, Gainesville, FL 32611}
\newcommand{\UCIrvine}{Department of Physics and Astronomy, University of California, Irvine, Irvine, California 92697-4575, USA}
\newcommand{\CBPF}{Centro Brasileiro de Pesquisas F\'{i}sicas, Rua Dr. Xavier Sigaud 150, Urca, Rio de Janeiro, Rio de Janeiro, 22290-180, Brazil}
\newcommand{\PUCP}{Secci\'{o}n F\'{i}sica, Departamento de Ciencias, Pontificia Universidad Cat\'{o}lica del Per\'{u}, Apartado 1761, Lima, Per\'{u}}
\newcommand{\INRM}{Institute for Nuclear Research of the Russian Academy of Sciences, 117312 Moscow, Russia}
\newcommand{\Jlab}{Jefferson Lab, 12000 Jefferson Avenue, Newport News, VA 23606, USA}
\newcommand{\Pittsburgh}{Department of Physics and Astronomy, University of Pittsburgh, Pittsburgh, Pennsylvania 15260, USA}
\newcommand{\Guanajuato}{Campus Le\'{o}n y Campus Guanajuato, Universidad de Guanajuato, Lascurain de Retana No. 5, Colonia Centro, Guanajuato 36000, Guanajuato M\'{e}xico.}
\newcommand{\Athens}{Department of Physics, University of Athens, GR-15771 Athens, Greece}
\newcommand{\Tufts}{Physics Department, Tufts University, Medford, Massachusetts 02155, USA}
\newcommand{\WM}{Department of Physics, College of William \& Mary, Williamsburg, Virginia 23187, USA}
\newcommand{\FNAL}{Fermi National Accelerator Laboratory, Batavia, Illinois 60510, USA}
\newcommand{\Purdue}{Department of Chemistry and Physics, Purdue University Calumet, Hammond, Indiana 46323, USA}
\newcommand{\MCLA}{Massachusetts College of Liberal Arts, 375 Church Street, North Adams, MA 01247}
\newcommand{\UMD}{Department of Physics, University of Minnesota -- Duluth, Duluth, Minnesota 55812, USA}
\newcommand{\Northwestern}{Northwestern University, Evanston, Illinois 60208}
\newcommand{\UNI}{Universidad Nacional de Ingenier\'{i}a, Apartado 31139, Lima, Per\'{u}}
\newcommand{\Rochester}{University of Rochester, Rochester, New York 14627 USA}
\newcommand{\Austin}{Department of Physics, University of Texas, 1 University Station, Austin, Texas 78712, USA}
\newcommand{\USM}{Departamento de F\'{i}sica, Universidad T\'{e}cnica Federico Santa Mar\'{i}a, Avenida Espa\~{n}a 1680 Casilla 110-V, Valpara\'{i}so, Chile}
\newcommand{\Geneva}{University of Geneva, 1211 Geneva 4, Switzerland}
\newcommand{\Chicago}{Enrico Fermi Institute, University of Chicago, Chicago, IL 60637 USA}
\newcommand{\hired}{}
\newcommand{\OregonState}{Department of Physics, Oregon State University, Corvallis, Oregon 97331, USA}
\newcommand{\bmeThanks}{now at SLAC National Accelerator Laboratory, Stanford, California 94309 USA}
\newcommand{\higueraThanks}{University of Houston, Houston, Texas, 77204, USA}
\newcommand{\damartinezThanks}{Now at Illinois Institute of Technology}
\newcommand{\joelmousseauThanks}{now at University of Michigan, Ann Arbor, MI, 48109}
\newcommand{\twaltonThanks}{now at Fermi National Accelerator Laboratory, Batavia, IL USA 60510}
\newcommand{\jwolcottThanks}{Now at Tufts University, Medford, Massachusetts 02155, USA }

% 60 total signatories.

\author{P.A.~Rodrigues}                   \affiliation{\Rochester}
\author{J.~Demgen}                          \affiliation{\UMD}
\author{E.~Miltenberger}                          \affiliation{\UMD}
\author{L.~Aliaga}                        \affiliation{\WM}
\author{O.~Altinok}                       \affiliation{\Tufts}
\author{L.~Bellantoni}                    \affiliation{\FNAL}
\author{A.~Bercellie}                     \affiliation{\Rochester}
\author{M.~Betancourt}                    \affiliation{\FNAL}
\author{A.~Bodek}                         \affiliation{\Rochester}
\author{A.~Bravar}                        \affiliation{\Geneva}
\author{H.~Budd}                          \affiliation{\Rochester}
\author{T.~Cai}                          \affiliation{\Rochester}
\author{M.F.~Carneiro}                    \affiliation{\CBPF}
\author{J.~Chvojka}                       \affiliation{\Rochester}
\author{J.~Devan}                         \affiliation{\WM}
\author{S.A.~Dytman}                      \affiliation{\Pittsburgh}
\author{G.A.~D\'{i}az~}                   \affiliation{\Rochester}  \affiliation{\PUCP}
\author{B.~Eberly}\thanks{\bmeThanks}     \affiliation{\Pittsburgh}
\author{M.~Elkins}                     \affiliation{\UMD}
\author{J.~Felix}                         \affiliation{\Guanajuato}
\author{L.~Fields}                        \affiliation{\FNAL}  \affiliation{\Northwestern}
\author{R.~Fine}                          \affiliation{\Rochester}
\author{A.M.~Gago}                        \affiliation{\PUCP}
\author{R.~Galindo}                        \affiliation{\USM}
\author{H.~Gallagher}                     \affiliation{\Tufts}
\author{A.~Ghosh}                         \affiliation{\CBPF}  \affiliation{\Rochester}
\author{T.~Golan}                         \affiliation{\Rochester}  \affiliation{\FNAL}
\author{R.~Gran}                          \affiliation{\UMD}
\author{D.A.~Harris}                      \affiliation{\FNAL}
\author{A.~Higuera}\thanks{\higueraThanks}  \affiliation{\Rochester}  \affiliation{\Guanajuato}
\author{K.~Hurtado}                       \affiliation{\CBPF}  \affiliation{\UNI}
\author{M.~Kiveni}                        \affiliation{\FNAL}
\author{J.~Kleykamp}                      \affiliation{\Rochester}
\author{M.~Kordosky}                      \affiliation{\WM}
\author{T.~Le}                            \affiliation{\Tufts}  \affiliation{\Rutgers}
\author{J.R.~Leistico}                          \affiliation{\UMD}
\author{A.~Lovlein}                          \affiliation{\UMD}
\author{E.~Maher}                         \affiliation{\MCLA}
\author{S.~Manly}                         \affiliation{\Rochester}
\author{W.A.~Mann}                        \affiliation{\Tufts}
\author{C.M.~Marshall}                    \affiliation{\Rochester}
\author{D.A.~Martinez~Caicedo}\thanks{\damartinezThanks}  \affiliation{\FNAL}
\author{K.S.~McFarland}                   \affiliation{\Rochester}  \affiliation{\FNAL}
\author{C.L.~McGivern}                    \affiliation{\Pittsburgh}
\author{A.M.~McGowan}                     \affiliation{\Rochester}
\author{B.~Messerly}                      \affiliation{\Pittsburgh}
\author{J.~Miller}                        \affiliation{\USM}
\author{A.~Mislivec}                      \affiliation{\Rochester}
\author{J.G.~Morf\'{i}n}                  \affiliation{\FNAL}
\author{J.~Mousseau}\thanks{\joelmousseauThanks}  \affiliation{\Florida}
\author{T.~Muhlbeier}                     \affiliation{\CBPF}
\author{D.~Naples}                        \affiliation{\Pittsburgh}
\author{J.K.~Nelson}                      \affiliation{\WM}
\author{A.~Norrick}                       \affiliation{\WM}
\author{Nuruzzaman}         \affiliation{\Rutgers}\affiliation{\Hampton}
\author{J.~Osta}                      \affiliation{\FNAL} 
\author{V.~Paolone}                       \affiliation{\Pittsburgh}
\author{C.E.~Patrick}                     \affiliation{\Northwestern}
\author{G.N.~Perdue}                      \affiliation{\FNAL}  \affiliation{\Rochester}
\author{M.A.~Ramirez}                     \affiliation{\Guanajuato}
\author{R.D.~Ransome}                     \affiliation{\Rutgers}
\author{H.~Ray}                           \affiliation{\Florida}
\author{L.~Ren}                           \affiliation{\Pittsburgh}
\author{D.~Rimal}                         \affiliation{\Florida}
\author{D.~Ruterbories}                   \affiliation{\Rochester}
\author{H.~Schellman}                     \affiliation{\OregonState}  \affiliation{\Northwestern}
\author{D.W.~Schmitz}                     \affiliation{\Chicago}  \affiliation{\FNAL}
\author{C.J.~Solano~Salinas}              \affiliation{\UNI}
\author{N.~Tagg}                          \affiliation{\Otterbein}
\author{B.G.~Tice}                        \affiliation{\Rutgers}
\author{E.~Valencia}                      \affiliation{\Guanajuato}
\author{T.~Walton}\thanks{\twaltonThanks}  \affiliation{\Hampton}
\author{J.~Wolcott}        \affiliation{\Tufts} \affiliation{\Rochester}
\author{M.~Wospakrik}                      \affiliation{\Florida}
\author{G.~Zavala}\thanks{\deceased}       \affiliation{\Guanajuato}
\author{D.~Zhang}                         \affiliation{\WM}

%
%% END AUTOMATIC PART
\collaboration{\minerva  Collaboration}\ \noaffiliation

\date{\today}

\begin{abstract}

Two different nuclear-medium effects are isolated using a low three-momentum transfer subsample of neutrino-carbon scattering data from the MINERvA neutrino experiment.
The observed hadronic energy in charged-current $\nu_\mu$ interactions is combined with muon kinematics to permit separation of the quasielastic and $\Delta$(1232) resonance processes.
First, we observe a small cross section at very low energy transfer that matches the expected screening effect of long-range nucleon correlations.  Second, additions to the event rate in the kinematic region between the quasielastic and $\Delta$ resonance processes are needed to describe the data.  The data in this kinematic region also has an enhanced population of multi-proton final states.  Contributions predicted for scattering from a nucleon pair have both properties;  the model tested in this analysis is a significant improvement but does not fully describe the data.  
We present the results as a double-differential cross section to enable further investigation of nuclear models.  Improved description of the effects of the nuclear environment are required by current and future neutrino oscillation experiments.

\end{abstract}
\ifnum\sizecheck=0
\maketitle
\fi

%\section{introduction}
The environment of the nucleus modifies neutrino-scattering cross sections, compared to those for hydrogen and deuterium targets.   
Fermi-gas models \cite{Smith:1972xh} 
% first citation in the paper !   energy secretary.  duh.
are still widely used by neutrino experiments to describe the nuclear environment, but incorporate only simple properties such as Fermi motion and Pauli blocking.  Such models are unable to precisely describe high-statistics data for neutrino scattering from oxygen \cite{Gran:2006jn}, carbon \cite{Lyubushkin:2008pe, Espinal:2007, AguilarArevalo:2010zc, Fiorentini:2013ezn, Walton:2014esl, Eberly:2014mra}, and iron \cite{Adamson:2014pgc}, especially for processes at low three-momentum transfer such as quasielastic (QE) and $\Delta$(1232) resonance production. The prevailing interpretation of these discrepancies is that more detailed nuclear models are required \cite{Gallagher:2011zza,Formaggio:2013kya,Alvarez-Ruso:2014bla}.  Uncertainties in nuclear modeling also impede investigation into fundamental quantities like the nucleon axial form factor.

The measurement of neutrino oscillation parameters by current and
future accelerator-based experiments  \cite{Adamson:2013ue,
  Abe:2015awa, Adamson:2016xxw, Adamson:2016tbq, Acciari:2015zzz, Adams:2013qkq,Acciarri:2015uup} requires accurate prediction of the neutrino energy spectrum.   Poorly modeled nuclear effects for the QE and $\Delta$ processes, or absence of an entire process 
such as interactions with correlated nucleon pairs, are major barriers \cite{Nieves:2012yz,Lalakulich:2012hs,Martini:2012uc,Mosel:2013fxa,Coloma:2013tba,Ericson:2015cva}  for these experiments.
The consequences are acute when the lepton kinematics or hadron final-state content bias neutrino energy reconstruction or might affect neutrinos and anti-neutrinos differently.   The data presented in this Letter reveal the magnitude of multi-nucleon effects that must be accounted for in oscillation parameter measurements.

We present the first analysis of neutrino-scattering data to isolate the kinematic region between  the QE and $\Delta$ resonance processes.  We use the measured hadronic energy to determine the full kinematics of an inclusive sample of interactions. 
These data from the MINERvA experiment exhibit a process with multiple protons in the final state, such as those  predicted by scattering from two particles leaving two holes (2p2h), 
with energy transfer between the QE and $\Delta$ reactions \cite{Martini:2009uj, Nieves:2011pp}.
Also, the cross section at low energy transfer is small, consistent with the effects of long range nucleon-nucleon correlations, such as those computed using the Random Phase Approximation (RPA) technique \cite{Nieves:2004wx, Martini:2009uj,Pandey:2014tza}.  
In this Letter, we first present the analysis strategy and hadronic energy estimators, then the selection of the low three-momentum sample, comparison of the data and simulated events in reconstructed quantities, and the extraction of a double-differential cross section which will enable further comparisons to interaction models.  The presence of a multi-nucleon component is confirmed by directly counting protons near the neutrino interaction point. 

%%%%%%%%%%%%%%%%%%%%%%%%%%%%%%%%%%%%%%%%%%%%%%%%%%%%%%%%%%%%%%
% Rik's version
%%%%%%%%%%%%%%%%%%%%%%%%%%%%%%%%%%%%%%%%%%%%%%%%%%%%%%%%%%%%%%
A productive approach in previous investigations of nuclear effects in
neutrino scattering has been to select a sample of QE events, measure
the final-state charged lepton kinematics, use them to infer $Q^2$
(the square of the four-momentum transferred to the nucleus), then
compare to models.  Predicted RPA and 2p2h effects overlap in $Q^2$,
despite distinctly different energy and momentum transfers \cite{Gran:2013kda}.  
Without a mono-energetic neutrino beam
or detailed convolution with the flux, these nuclear medium effects
are difficult to distinguish using only muon kinematics.

%%%%%%%%%%%%%%%%%%%%%%%%%%%%%%%%%%%%%%%%%%%%%%%%%%%%%%%%%%%%%%
% Phil's version
%%%%%%%%%%%%%%%%%%%%%%%%%%%%%%%%%%%%%%%%%%%%%%%%%%%%%%%%%%%%%%
%Previous investigations of nuclear effects in neutrino scattering have
%typically selected a sample of QE events, measured the final-state
%lepton kinematics, and used them to infer $Q^2$ (the square of the
%four-momentum transferred to the nucleus). RPA and 2p2h effects
%overlap in $Q^2$, despite distinctly different energy and momentum
%transfers~\cite{Gran:2013kda}, making them difficult to distinguish
%using only muon kinematics.

Reconstructing both hadronic energy and muon kinematics permits an estimate of the neutrino energy $E_\nu$ plus an electron-scattering style analysis of a pair of variables which separate QE and $\Delta$ events.  This pair can be either hadronic invariant mass $W$ and $Q^2$, or energy transfer $q_0$ and the magnitude of three-momentum transfer $q_3=|\vec{q}|$ to the nucleus.  The latter basis is used in this analysis, to avoid a model dependence inherent in producing an unfolded cross section for regions of $W$ where the default model predicts almost no cross section.  

Reconstruction of the energy transfer $q_0$ requires model-dependent corrections for nucleon removal energy and unobserved neutrons.  Additionally, the QE and $\Delta$ processes contribute zero cross section at some kinematics, which prevents unfolding to true $q_0$. 
To produce a double-differential cross section with little model dependence,
we define the closely related observable, the hadronic energy available to produce activity in the detector \Eavail , as the sum of proton and charged pion kinetic energy, plus neutral pion, electron, and photon total energy, and report \dsdEdq. The precision of \Eavail depends primarily on the accurate simulation of charged particles and photons that leave the interaction point and deposit energy throughout the detector.

These data are taken from the 2010 to 2012 MINERvA exposure to the NuMI beam with $3.33 \times 10^{20}$ protons on target.  In the neutrino-mode NuMI beam, 120-GeV protons interact with a graphite target, and positively-charged mesons are focused toward the MINERvA detector by a pair of magnetic horns \cite{Adamson:2015dkw}.  The mesons decay to neutrinos in a helium-filled decay pipe, leading to a neutrino event spectrum which %for this low-energy tune which %for these data
peaks at 3.5~GeV.
The neutrino flux prediction comes from a Geant4-based \cite{Agostinelli2003250,1610988} simulation of the neutrino beamline, tuned using thin-target hadron production data~\cite{Alt:2006fr, Denisov:1973zv, Carroll:1978hc,Allaby:1969de} with additional uncertainty assigned to interactions not constrained by those data. 

An inclusive sample of $\nu_\mu$ charged-current interactions is selected using events that originate in MINERvA's 5.3-ton %(3.17x10$^{30}$ nucleons), 
active-tracker fiducial volume~\cite{Aliaga:2013uqz}, which consists of planes of triangular scintillator strips with a 3.4-cm base and 1.7-cm height which are up to 2~m long.  Hydrogen, carbon, and oxygen account for 7.4\%, 88\%, and 3.2\% of the target nuclei by weight.  The planes are hexagonal and alternate between three orientations ($0^{\circ}$ and $\pm 60^{\circ}$)  around the detector axis, enabling a precise reconstruction of the interaction point and muon track angle, even when hadronic activity overlaps the muon in one orientation. Muon tracks in MINERvA are matched to tracks in the MINOS near detector~\cite{Michael:2008bc}, a magnetized iron spectrometer located 2~m downstream of \minerva that measures the muon momentum and charge of the muons.

The muon energy is measured from its range if the muon stops in MINOS, otherwise by curvature in the MINOS magnetic field.  That energy in MINOS is added to an estimate from the muon's range in \minerva to form $E_\mu$ and $p_\mu$, the muon energy and momentum.  The muon angle $\theta_\mu$ is measured by tracking the muon in MINERvA from the interaction point.  To produce an unfolded cross section based on data in regions with good muon acceptance, we require $\theta_\mu<20^\circ$ and $E_\mu>1.5$~GeV for the selection and later when unfolding. 

The hadronic energy is reconstructed from the summed energy in the MINERvA detector not associated with the muon.    A Monte Carlo (MC) simulation, based on the {\sc genie} neutrino interaction generator \cite{Andreopoulos201087} and a Geant4 simulation of the detector, is used to obtain corrections of this summed energy to both \Eavail and $q_0$.  The latter correction depends significantly on the neutrino interaction model, especially the predicted neutron content of the final state.  
The rest of the kinematics are neutrino energy $E_\nu = E_\mu+q_0$,  from which we form the square of the four-momentum transfer
$Q^2 = 2E_\nu(E_\mu - p_\mu\cos\theta_\mu) - M^2_\mu$,
($M_\mu$ is the muon mass) and the three-momentum transfer $q_3 = \sqrt{Q^2 + q_0^2}$.  Results are presented in slices of $q_3$, which includes a $q_0$ model dependence diluted by muon energy and angle contributions.  The resolution of $q_3$ is 22\%, dominated by the resolution of $q_0$.  

The event selection is completed by requiring $2 < E_\nu < 6$~GeV, an interval chosen to span the peak of the neutrino flux.   A subsample is formed into six bins of $q_3$ from 0~to~0.8~GeV for presentation of the cross section, which for brevity are combined into two ranges of $q_3$ 
when showing data with reconstructed kinematics.  There are 74,749 events in this data sample with average $E_\nu$ of 3.9~GeV.

We estimate the reconstructed \Eavail using just the calorimetric sum of energy (not associated with the muon) in the central tracker region and the electromagnetic calorimeter region immediately downstream of the tracker. The unrelated beam activity from data is overlaid directly onto simulated events.  The outer tracking and calorimetric regions of the MINERvA detector are not included; they contain activity from neutrons and photons, but they capture more unrelated beam activity which biases \Eavail .  The resulting \Eavail resolution varies from 55\% to 38\% for $q_3$ from 0 to 0.8~GeV.   The neutron content of the {\sc genie} model plays a minor role via the \Eavail resolution.  Test beam constraints on calorimetry and Birks' suppression for MINERvA scintillator are used to tune the simulation and set the uncertainty on the single-particle response \cite{Aliaga:2015aqe} .  The detector's simulated calorimetric response to protons and pions typical of the low-$q_3$ sample agrees with data from the MINERvA hadron test beam experiment.

The neutrino interaction model is from {\small GENIE 2.8.4}.  The QE model uses a relativistic Fermi gas with an axial mass parameter of 0.99 GeV.  The resonance production is from Rein-Sehgal \cite{Rein:1980wg} with a {\sc genie}-specific non-resonant background, and a transition to deep inelastic scattering from $W>1.7$~GeV.  Events with a pion in the final state are part of this inclusive charged-current selection, and have been shown in previous \minerva analyses~\cite{Higuera:2014azj,Eberly:2014mra} to be overestimated by {\sc genie}.
We use those results to  modify the prediction: the one-pion neutrino-neutron non-resonant component is reduced by 75\% \cite{Wilkinson:2014yfa,Wilkinson:2015zzz}, and the total rate of pion production with $W<1.8$~GeV is further reduced by 10\%. Coherent pion production with $E_\pi < 450$~MeV is also reduced by 50\%.  We refer to this tuned simulation as the default model in this Letter.  

\begin{figure}[t]
\begin{center}
\includegraphics[width=8.7cm]{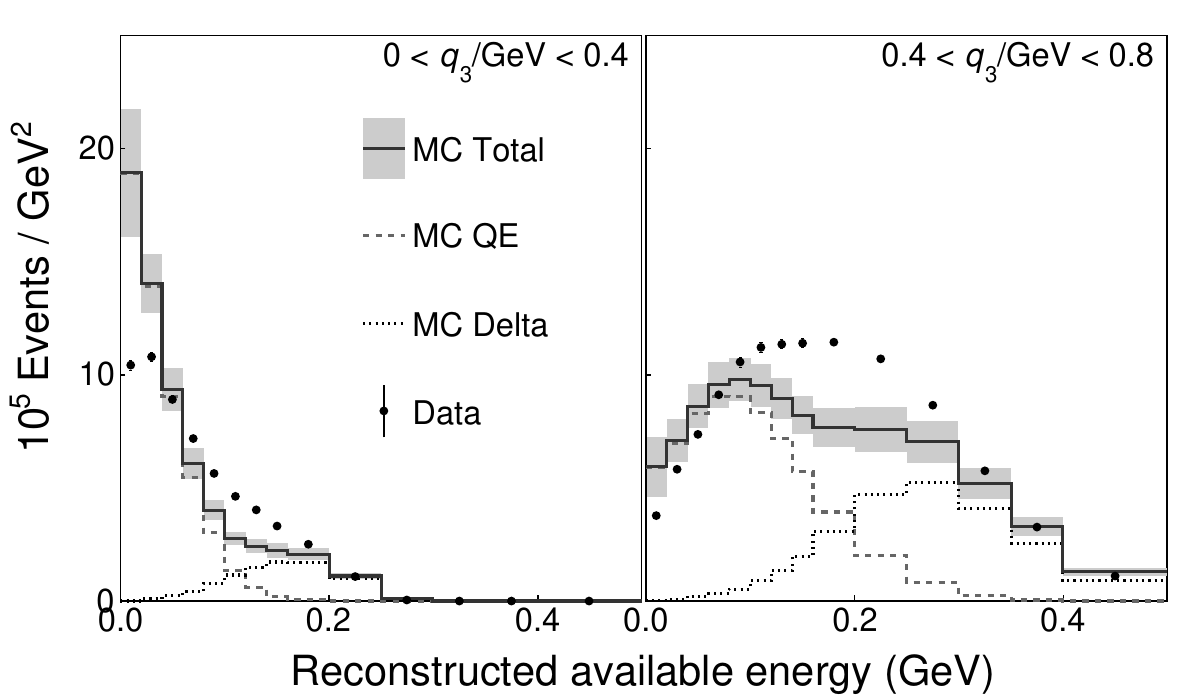}
\caption{Reconstructed \Eavail compared to the default simulation for two ranges of reconstructed three momentum transfer.
%, $q_3 < 0.4$~GeV (left) and $0.4 < q_3 < 0.8$~GeV (right).  
The region between the predicted QE process (dashed line) and the $\Delta$(1232) resonance (dotted line) is filled in by an unmodeled process.  The lowest \Eavail data is far below the simulation.  Data are shown with statistical uncertainty only, which is too small to see.  The absolutely normalized simulation is shown with systematic uncertainties.
\label{fig:reco}}
\end{center}
\end{figure}

The distribution of reconstructed \Eavail is shown in Fig.~\ref{fig:reco} and compared to the simulation.
Both halves of the $q_3$ range show the same discrepancies: the simulation has too many QE events and too few events in the region between the QE and $\Delta$ processes.

To study detailed effects of the nucleus, we construct additional comparisons by modifying {\sc genie}'s description of the quasielastic process with the RPA effect from the calculation of Nieves~\emph{et al.}~\cite{Nieves:2004wx}.  A two-dimensional correction in $(q_0,q_3)$ is formed from the ratio of cross sections between the model with RPA effects and the model without, and applied to the {\sc genie} quasielastic cross section.  
The RPA model does include a short range correlation effect, but we do not simulate the presence of the spectator nucleon \cite{Egiyan:2005hs,Subedi:2008zz} in the final state.

We also add a 2p2h process for carbon and oxygen to the simulation,
using the IFIC Valencia model~\cite{Nieves:2011pp, Gran:2013kda}. The cross section depends on $q_0$, $q_3$, and whether the nucleon pair involved in the initial interaction was proton-neutron or neutron-neutron.  This calculation includes only the QE-like (no pion in the final state) contributions, not 2p2h1$\pi$ (with a pion).  It also includes interactions with $\Delta$ kinematics, but not higher-mass resonances. 

Explicit hadron kinematics are added to the 2p2h model using a strategy similar to that of Ref.~\cite{Sobczyk:2012ms}, documented in detail in Ref.~\cite{Gran:2015xyz}.  The nucleons have momenta drawn from the standard {\sc genie} Fermi gas distribution, and are given one unit charge and the momentum and energy transfer from the lepton, less 25~MeV removal energy for each nucleon.  The final momentum is distributed between the pair as in an isotropic decay in the center of momentum frame,  which is a good approximation~\cite{Simo:2014esa} to a full calculation.   The resulting nucleons are passed to the {\sc genie} intranuclear rescattering model where their number, angle, and energy may change.

An unfolding procedure~\cite{D'Agostini:1994zf} with four iterations is applied in two dimensions to translate the data from reconstructed quantities to true $(\Eavail,q_3)$.  
The simulation is used to correct for the acceptance of the fiducial volume, the efficiency of the MINOS muon match, and the subtraction of small (3\%) neutral-current and $\mu^+$ backgrounds.  Dividing by the flux and $3.17 \times 10^{30}$ nucleon targets results in the double-differential cross section \dsdEdq, shown\footnote{Tables of this cross section and the estimated flux are available in the supplementary material.}
in Fig.~\ref{fig:crosssection} for six ranges of $q_3$.  

%\begin{figure*}[t]
\begin{figure}[t]
\begin{center}
\includegraphics[width=8.7cm]{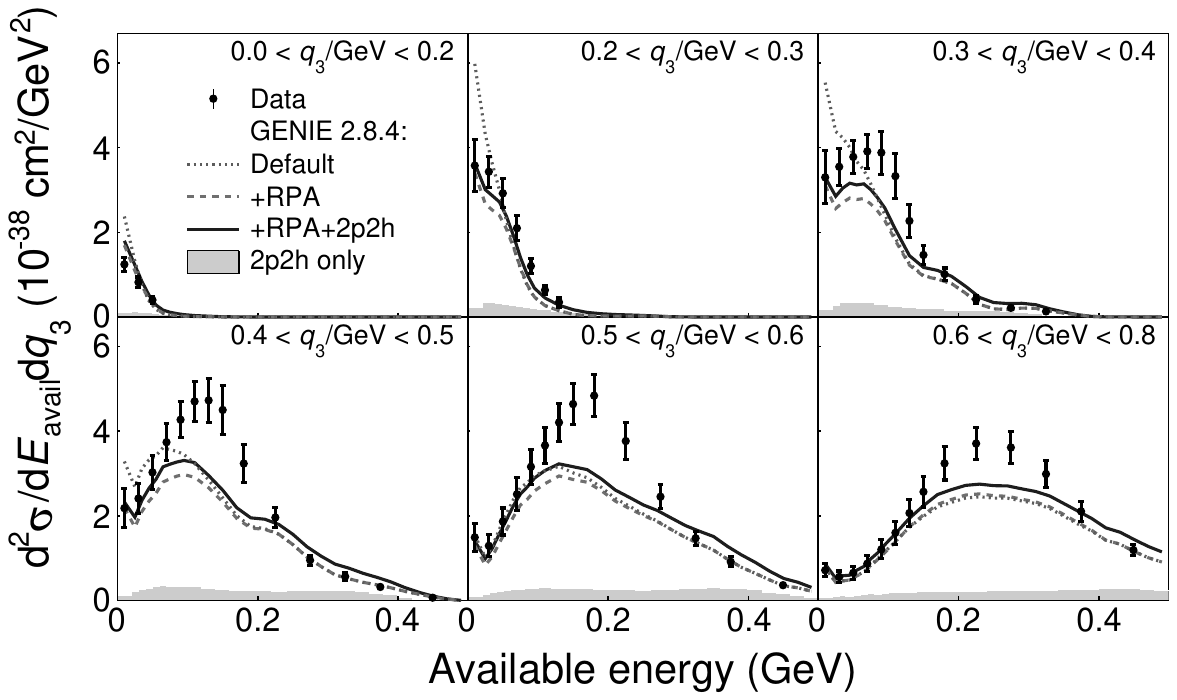}
\caption{The double-differential cross section \dsdEdq in six regions of $q_3$ is compared to the {\small GENIE 2.8.4} model with  reduced pion production (small dot line), the same with RPA suppression (long-dashed), and then combined with a QE-like 2p2h component (solid).  The 2p2h component is shown separately as a shaded region. {\sc genie} predicts events with zero available energy (all neutrons in the final state); as is done here in order to compare to data, the cross section must be summed including the spike at zero to the edge of the the first bin in each $q_3$ range to produce an average cross section.
\label{fig:crosssection}}
\end{center}
\end{figure}
%\end{figure*}

Both the $q_3$ and the \Eavail estimators have mild dependence on the interaction model.  The results in this Letter, especially the migration matrix used for the unfolding, are produced using the fully-modified model rather than the default model.   Since the fully-modified model does not provide a complete description of the data, we also extract the cross section using the default model, and take the difference as a systematic uncertainty.  This is the largest contributor (10\%) to the systematic uncertainty for $q_3$ below 0.4~GeV.  The flux uncertainty (9\%) is the next largest, followed by hadronic and muon energy scales.  The total uncertainty ranges from 10\% at high $q_3$ and high \Eavail, growing to 20\% at the lowest \Eavail and $q_3$.

The discrepancy seen in the unfolded data in Fig.~\ref{fig:crosssection} is much smaller with these model additions.  The  RPA suppression has a significant effect on the lowest $E_{\mathrm{avail}}$ bins, and produces very good agreement.  The RPA model is theoretically motivated and the lowest $Q^2$ behavior is tuned to external data,  neutron decay for the axial form factor $F_A(Q^2=0)$, and muon capture on nuclei~\cite{Nieves:2004wx} for the long-range correlation effect.  The $\chi^2$ from comparing the simulation to reconstructed data, with the full covariance matrix and six bins of $q_3$, decreases from 896 (for 61 degrees of freedom) for the default simulation to 540 when the RPA effects are added.
The simulated QE-like 2p2h contribution spans the horizontal axis and mitigates some of the discrepancy in the region between the QE and $\Delta$.
The resulting $\chi^2$ is improved further to 498, but this prediction still does not fully describe the data.

The unmodeled shape differences between the data and models shown in Fig.~\ref{fig:reco} are the same (within statistical uncertainties) as samples from a higher energy range $6 < E_\nu < 20$~GeV selected from the same run period.
Differences in the normalization of high and low energy samples are consistent with the energy-dependent uncertainties of the flux.   An extreme case of zero 2p2h component above 5~GeV is disfavored by more than three standard deviations, with the muon energy scale being the largest systematic uncertainty. This favors the hypothesis that the apparent tension between MiniBooNE~\cite{AguilarArevalo:2010zc} and NOMAD~\cite{Lyubushkin:2008pe} arises from differences in selecting multi-proton final states, and not from strong neutrino energy dependent nuclear effects.
The lack of energy dependence is also confirmation that the low-$\nu$ method \cite{Mishra:1990ax, seligman, Adamson:2009ju, Bodek:2012uu,Josh-thesis} 
may be effective in constraining the relative $E_\nu$ dependence of the neutrino flux, even with unmodeled nuclear effects.

%%%%%%%%%%% proton counting

There is an independent marker for a multi-nucleon component; the 2p2h process transfers energy and momentum to two nucleons, which will be ejected from the nucleus.  This is in contrast to the QE, $\Delta$, and coherent pion interactions which produce a single recoiling nucleon, nucleon and pion, and only a pion, respectively, before final state interactions (FSI).  
The IFIC Valencia model predicts \cite{Gran:2013kda} that proton plus neutron initial states are 50 to 80\% of the total.   The presence of additional protons was inferred from the energy spectrum of hadronic activity near the neutrino interaction point of QE events in an earlier \minerva result~\cite{Fiorentini:2013ezn}.  Another observation of proton pairs is reported by ArgoNeuT~\cite{Acciarri:2014gev}.  Using a technique to effectively count protons, we find the data have more events with two or more observable protons in the final state, compared to the default model. 

This analysis identifies protons in \minerva directly using the Bragg peak at the end of their range in scintillator: protons are likely to deposit 20~MeV or more in the scintillator strip where they stop (which may be the strip where the interaction occurred). We define a search region around the neutrino interaction point $\pm 170$~mm in the beam direction and $\pm 83$~mm in the transverse direction. Pions and neutrons are likely to exit the search region or leave only low energy deposits. Simulated QE and $\Delta$ production events with a pion
from the $0.4 < q_3 < 0.8$ sample produce an average of 1.0 strips with activity more than 20~MeV in the interaction region.  Two-nucleon events from the 2p2h process or $\Delta$ interactions that lose their pion to FSI produce an average of 1.6 and 1.5 strips with 20~MeV, respectively.

%%% Attempt to make a figure

\begin{figure}[t]
\begin{center}
\includegraphics[width=4.2cm]{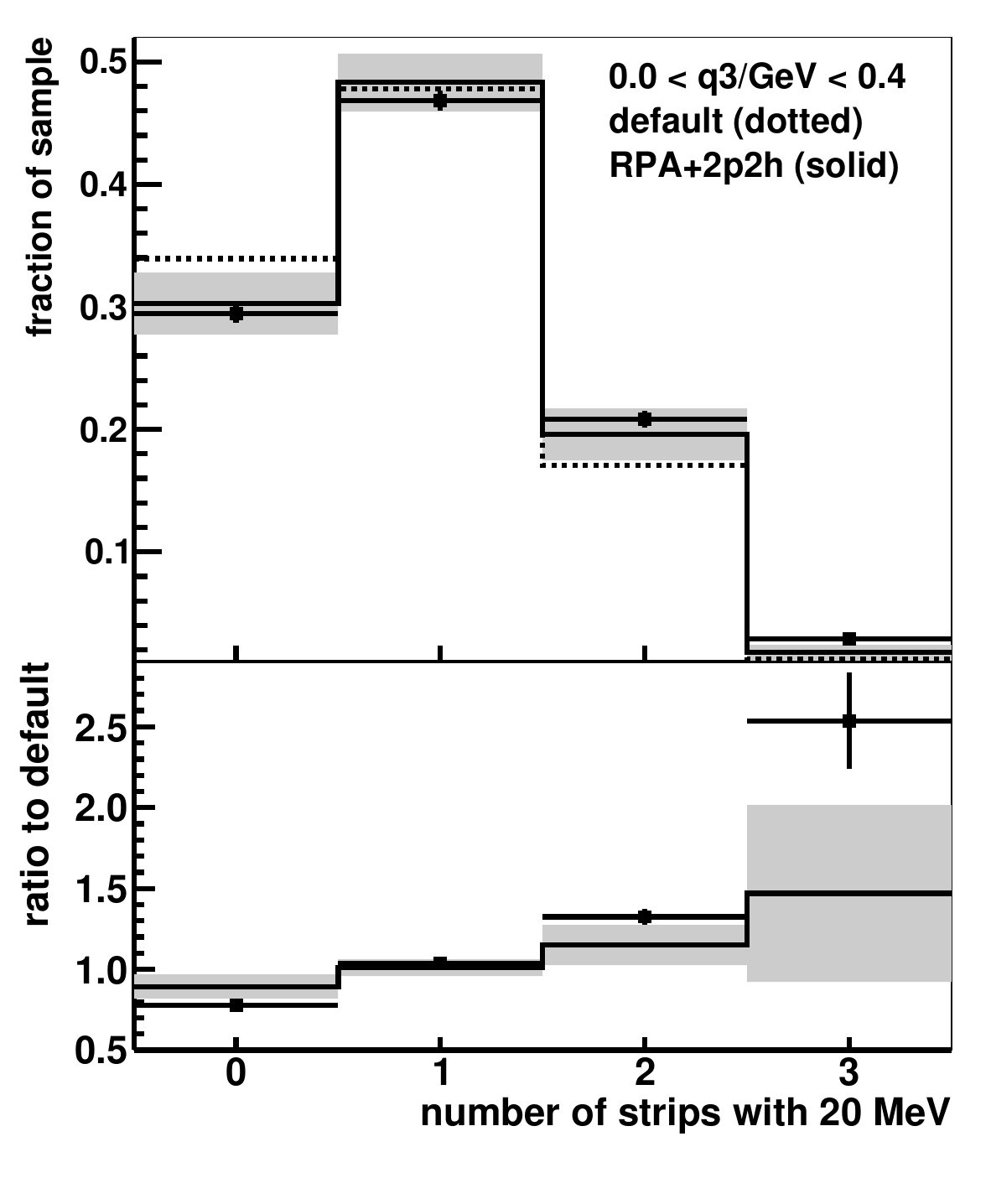}
\includegraphics[width=4.2cm]{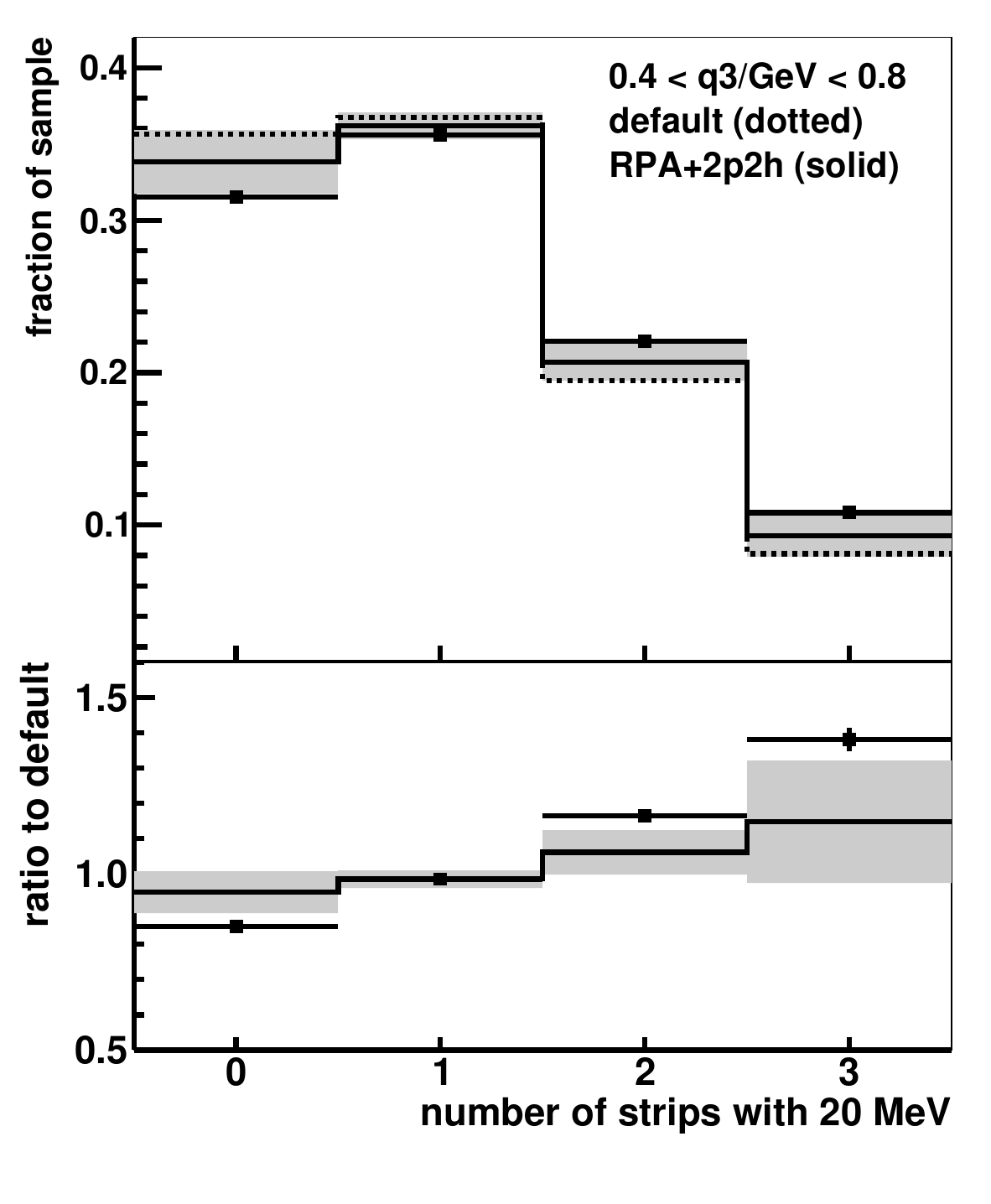}
\caption{Fraction of events with zero, one, two, and three or more
  strips with at least 20 MeV of activity near the interaction point,
  The samples are from the region between QE and $\Delta$ 
  for  two ranges of reconstructed three momentum transfer.
  %, $q_3 < 0.4$~GeV (left) and $0.4 < q_3 < 0.8$~GeV (right).  
  The model with RPA and 2p2h is shown with the solid line and systematic uncertainty band; the data are shown with statistical uncertainties.  The ratios are taken with respect to the default model, shown as a dotted line. 
RPA suppression negligibly modifies the default model for this quantity and is not shown.
\label{fig:protoncounting}}
\end{center}
\end{figure}

The simulation has fewer protons than the data, and the 2p2h model simulated for this analysis is essential to obtain agreement.   Figure~\ref{fig:protoncounting} shows the multiplicity of deposits above $20$~MeV observed in data and variations of the model, for the region between the QE and $\Delta$ processes, specifically from 0.08 to 0.16~GeV (0.14 to 0.26~GeV) in the left (right) $q_3$ distributions from Fig.~\ref{fig:reco}.  The addition of the 2p2h component makes the most dramatic change.  The combined $\chi^2$ improves 15.1 to 7.5 for six degrees of freedom.  More multi-nucleon events would further improve agreement. 
The model with RPA and {\small GENIE 2.8.4} model without reduced pion production (neither are shown) yields a $\chi^2$ of 15.2 and 19.6, respectively.

%%%%  End attempt to make a figure

The region at higher \Eavail, dominated by resonances and with unsimulated 2p2h1$\pi$ interactions, shows all the same trends.  In the QE region at lower \Eavail, the agreement is most improved with the addition of the RPA suppression; sensitivity to multiple protons is reduced due to the QE background and the protons' lower energy.

The most significant systematic uncertainty for the proton counting study is from the value~\cite{Aliaga:2015aqe} for Birks' parameter used in the detector simulation, though it plays a minor role in Fig.~\ref{fig:crosssection}.  Uncertainties from the FSI model, especially pion absorption, change the multi-nucleon content and are also significant, but the 1$\sigma$ uncertainty produces effects that are a factor of three smaller than this model for 2p2h reactions.  The shape of the pion energy spectrum reported in \cite{Eberly:2014mra} is especially sensitive to the FSI model and is adequately described with {\sc genie} and its FSI uncertainties.  

%%%%%%%%% Conclusions

The significantly improved agreement, even using a single 2p2h model with a simplified hadronic system, is additional evidence that a multi-nucleon component is present in the data.  
Refinements to this 2p2h model, or other models \cite{Martini:2009uj,  Megias:2014qva} not currently available for full simulation, may predict more multi-proton events or events with different kinematics, and further improve the description of the observed event rate and proton content of these samples.  Augmented treatment of the 1p1h, short-range correlation component, with constraints from the superscaling method \cite{Amaro:2004bs, Megias:2014qva} or a simulated final state that includes the spectator nucleon \cite{Bodek:2014pka}, may also contribute to a better simulation.

The data make clear two distinct multi-nucleon effects that are essential for complete modeling of neutrino interactions at low momentum transfer.  The 2p2h model tested in this analysis improves the description of the event rate in the region between QE and $\Delta$ peaks, and the rate for multi-proton events, but does not go far enough to fully describe the data.  Oscillation experiments sensitive to energy reconstruction effects from these events must account for this event rate.  The cross section presented here will lead to models with significantly improved accuracy.

%\ifnum\sizecheck=0
  \begin{acknowledgments}

We are grateful to the authors of the RPA and 2p2h models for making
the code for their calculations available for study and incorporation into this analysis.
This work was supported by the Fermi National Accelerator Laboratory
%, which is operated by the Fermi Research Alliance, LLC, 
under US Department of Energy contract
No. DE-AC02-07CH11359 which included the \minerva construction project.
Construction support was
also granted by the United States National Science Foundation under
Grant PHY-0619727 and by the University of Rochester. 
Support for scientists for this %% TAKE THIS OUT FOR FUTURE PAPERS!
specific publication was granted by the United States National Science
Foundation under Grant PHY-1306944.  Support for
participating scientists was provided by NSF and DOE (USA) by CAPES
and CNPq (Brazil), by CoNaCyT (Mexico), by CONICYT (Chile), by
CONCYTEC, DGI-PUCP and IDI/IGI-UNI (Peru), by Latin American Center for
Physics (CLAF) and by RAS and the Russian Ministry of Education and Science (Russia).  We
thank the MINOS Collaboration for use of its
near detector data. Finally, we thank the staff of
Fermilab for support of the beamline, the detector, and the computing infrastructure.

\end{acknowledgments}

  \bibliographystyle{apsrev4-1}
  \bibliography{rpamecpaper}

%\fi

%\ifnum\PRLsupp=0
%  \clearpage
%  \input{appendix.tex}
%\fi

\end{document}